\documentclass[conf]{IEEEtran}
\usepackage{graphicx}
\usepackage{amsmath}
\usepackage{amsthm}
\usepackage{bm}

\ifCLASSINFOpdf
 
\else
 
\fi

\begin{document}

\title{PhysioGait: Context-Aware Physiological Context Modeling for Person Re-identification Attack on Wearable Sensing}

\author{$^1$James OSullivan, $^{1,2}$Mohammad Arif Ul Alam,~\IEEEmembership{\\
$_1$University of Massachusetts, Lowell, MA\\
$_2$University of Massachusetts Chan Medical School, Worcester, MA\\
James\_OSullivan1@student.uml.edu, mohammadariful\_alam@uml.edu}
}

\maketitle

\begin{abstract}
Person re-identification is a critical privacy breach in publicly shared healthcare data. We investigate the possibility of a new type of privacy threat on publicly shared privacy insensitive large scale wearable sensing data. In this paper, we investigate user specific biometric signatures in terms of two contextual biometric traits, physiological (photoplethysmography and electrodermal activity) and physical (accelerometer) contexts. In this regard, we propose \emph{PhysioGait}, a context-aware physiological signal model that consists of a Multi-Modal Siamese Convolutional Neural Network (mmSNN) which learns the spatial and temporal information individually and performs sensor fusion in a Siamese cost with the objective of predicting a person's identity. We evaluated \emph{PhysioGait} attack model using 4 real-time collected datasets (3-data under IRB \#HP-00064387 and one publicly available data) and two combined datasets achieving 89\% - 93\% accuracy of re-identifying persons.
\end{abstract}

\begin{IEEEkeywords}
Siamese Network, Neural Network, Activity Recognition, Security and Privacy, Wearable sensing, Physiological Sensing
\end{IEEEkeywords}

\IEEEpeerreviewmaketitle

\section{Introduction}

In-situ physiological sensing also has been used in many places such as gymnasiums using touch activated Heart Rate (HR) estimation while running and doctors using wearable medical devices (ECG, Pressure meter) to estimate different physiological characteristics of humans during different activities. On the other hand, internet is flooded with de-identified Internet-of-Things data due to their privacy insensitive nature providing similar human contexts \cite{abd20,iot1}. However, experimental research findings indicate that many seemingly innocuous sensors can be exploited to infer highly sensitive information about people in their vicinity \cite{secure1}. In this paper, we investigated a potential privacy breach on health contexts extracted from wearable Accelerometer (ACC) (activity context) and physiological (photoplethysmography (PPG), Electrodermal Activity (EDA)) data.

Due to the immense necessity of monitoring Breathing Rate (BR) and Heart Rate (HR) continuously and efficiently, researchers developed efficient privacy insensitive commodity sensor technologies and sophisticated machine learning techniques for monitoring photoplethysmography (PPG), electrodermal activity (EDA), temperature using wearables. To progress research productivity, researchers often share raw sensor data and Protected Health Information (PHI) in public servers with appropriate Covered Entity (CE), encoding and de-identification according to privacy rule (such as Health Insurance Portability and Accountability Act (HIPAA) Privacy Rule in USA \cite{hipaa}). As per CE of HIPAA Privacy Rule \cite{hipaa}, raw PPG, EDA, Skin Temperature and ACC data are not privacy sensitive and do not require any encoding while sharing publicly. However, the recent evolution of face and remote PPG detection techniques using public camera recorded video data \cite{rppg1} or in-situ sensing (touch, radar, WiFi signal-based) raises an important privacy concern: {\it Does physiological sensing (PPG, EDA, Temp) coupled with inertial measurement units (accelerometer, gyroscope, magnetometer) carry any user specific biometric signature which can be utilized against privacy attack?}

Many researchers showed PPG can be utilized as an additional biometric signature along with user's face for personal device authentication purpose \cite{auth_ppg1} Though, face data has never been shared along with PHI wearable data, the user authentication method can not be used as an attack over user's privacy in publicly available HIPAA compliant wearable data. However, recently, ubiquitous computing researchers provide efficient quantification of physiological sensing (say HR) and contextual information (physical activities) to provide healthcare services including stroke prevention, epileptic seizure detection etc \cite{acc_ppg1,acc_ppg2}. This additional contextual information has been widely shared publicly as HIPAA privacy rule does not forbid sharing such data publicly. In this paper, we argue that, the quantification of simultaneously collected multi-modal sensor aided physical and physiological sensing carry unique biometric signatures of individuals which may involve a new type of potential privacy attack, a Person Re-identification Attack, which can be a serious privacy threat to billions of publicly shared healthcare data.


In this regard, we develop a multi-modal Siamese Neural Network (mmSNN) inspired context-aware physiological sensing (\emph{PhysioGait}) attack model, that provides the following {\bf key contributions}

\begin{itemize}
\item We develop an activity recognition and localization framework from wearable accelerometer data and develop an image representation of physical and physiological data within the activity window.

\item We develop a Multi-Modal Siamese Neural Network (mmSNN) that consists of two sub-neural-networks. mmSNN leverages a spatio-temporal architecture to generate multi-modal encodings of time series physiological sensing and image representation of physical and physiological data within the activity window.

\item We develop an efficient Siamese cost function to fuse time-series physiological sensor data and image representations of accelerometer data. We also develop an efficient weighted cost function on two sub-networks together as a decision function to find the matching of user's identity.

\item Finally, we evaluate our proposed \emph{PhysioGait} framework on 3 in-house collected data, one publicly available data set and two combined data sets.
\end{itemize}

\section{Related Works}
There are few works that proposed wearable motion sensors-based privacy attacks before \cite{wear_privacy1,wear_privacy2,wear_privacy3,wear_privacy4,wear_privacy5}. One of the first of its kind is BioInsights, which focused on wearable (watch and google glass) accelerometer and gyroscope based 3 body-posture (pre-post exercise) classification and personalized biometric signature identification using Support Vector Machine algorithm \cite{wear_privacy1}. Hussein et. al. presented an activity-independent person identification from wearables-based daily activity recognition using an Ensemble Bagged Trees algorithm \cite{wear_privacy4}. Kilic et. al. proposed to transform accelerometer, gyroscope and magnetometer sensor signals into image form for 19 different detected activities (sitting,standing, step exercise, leaping, basketball playing etc.) and utilized ResNet-based transfer learning to model biometric insights for identifying persons \cite{wear_privacy2}. Benegui et. al. proposed a gray-scale image representation of accelerometer, gyroscope and magnetometer data from wearable sensors and developed a Convolutional Neural Network (CNN) based few shot user identification technique \cite{wear_privacy3}. RLTIR proposed an interactive identity recognition approach based on reinforcement learning under human guidance which is the first attempt to combine human expert knowledge with model learning in the area of person re-identification from wearables \cite{wear_privacy5}.

Wearable physiological sensor signals (ECG, EEG) modeling also has been used for person identification before \cite{ecg_privacy1,ecg_privacy2,ecg_privacy3}. Lee et. al. proposed a robust principal component analysis network (RPCANet) and wavelet analysis technique to model user specific biometric signatures to identify people from a large scale ECG database \cite{ecg_privacy1}. Kim et. al. proposed bidirectional long short-term memory (LSTM)-based deep recurrent neural networks (DRNN) through late-fusion to develop a real-time system for ECG-based biometrics identification and classification \cite{ecg_privacy2}. Ko et. al. proposed a novel algorithm called adjusted (Q,S) that automatically balance the R-peak distribution of ECG signal towards modeling biometric signature for person re-identification from ECG data \cite{ecg_privacy3}. Apart from ECG or PPG signals (both are used for HR detection), EEG signals has been utilized for person re-identification attacks before as well \cite{eeg_privacy1}. There are no studies that have been performed investigating the Galvani Skin Response (GSR) i.e. Electrodermal Activity (EDA) based or spatiotemporal fusion of accelerometer and physiological signals for person re-identification before.

\emph{PhysioGait} is the first of its kind that combines human activities with their corresponding physiological responses modeling in miscroscoping level towards person re-identification. Taking advantage of the innovative contributions from existing wearable sensor signals processing approaches on accelerometer, PPG and EDA, \emph{PhysioGait} proposes a novel Multi-Modal Siamese Convolutional Neural Network (mmSNN) that learns preprocessed spatial and temporal sensor signals individually and performs sensor fusion in a Siamese cost with the objective of predicting a person's identity.

\section{Activity Feature Learning }


\subsection{Gesture Recognition and Localization}
We modify GestureKeeper proposed (\cite{gesture_keeper}) technique which is a robust hand-gesture recognition system based on a wearable inertial measurements unit (IMU) \cite{gesture_keeper}. In this regard, we first define 12 different hand gesture movements. GestureKeeper first identifies the start of a gesture recorded in IMU data stream using a recurrence quantification analysis (RQA) technique. RQA enables the detection of transitions in the dynamical behavior (e.g. deterministic, chaotic, etc.) of the observed system. A major advantage of RQA is its fully self-tuned nature, in the sense that no prior parameter fine-tuning is required in a manual fashion. We select a window size of 80 samples with 80\% overlap between consecutive windows, which represents approximately 2.5 seconds of IMU data (32 Hz frequency), sufficient for capturing even the longest of the dictionary's gestures. After detecting the starting point of hand gesture, we detect the gesture which is a 12 class problem. We use support vector machine (SVM) classifier using the radial kernel. As input of the SVM, we use both statistical features and samples of the acceleration signal features. The statistical features include the mean, median, root mean square (RMS), standard deviation, variance, skewness and kurtosis, of the 3D acceleration, angular velocity, and magnetism time series provided by the sensor. The sample based features are formed by a re-sampling process of the x, y and z axis acceleration. After re-sampling, the new signal is composed of a fixed number of samples for each acceleration time series. The final set of features includes the statistical ones (introduced earlier) along with this number of samples of the re-sampled acceleration signal for each of the x-, y-, and z-axis time series. Overall, we first utilize RQA in order to extract features for the identification sub-system, thereafter uses an SVM model, trained with the aforementioned features, for identifying correctly the windows that contain gestures providing the recognized gesture label (one of 12) along with the start and end point (localization) of it.

\subsection{Electrodermal Activity Decomposition into Tonic and Phasic components}
Any activity related to changes in the electrical characteristics of the epidermis is known as the EDA. Skin conductance (SC), a measure of EDA, is defined as the conductance measured across two distinct points in the epidermis in the presence of salty sweat secretions produced by different eccrine sweat glands \cite{ans5}. Depending on the physiological need, the autonomic nervous system (ANS) stimulates sweat glands to produce sweat which changes the conductivity of the epidermis \cite{ans5}. These changes in conductivity can be analyzed to understand the underlying ANS activity, especially the sympathetic nervous activity, which contains a great deal of information about human arousal. SC can be modeled as a summation of two components \cite{ans5}. The relatively slow varying component, which is called the tonic component, is generally dependent on the thermoregulation of the body, ambient temperature, and humidity. This tonic component also contains information about the general arousal of a person \cite{ans5}. On the other hand, the comparatively fast varying component, which is called the phasic component, is a reflection of neural stimulation from the sympathetic nervous system, a part of the autonomic nervous system (ANS) \cite{ans5}. Hence, EDA can be represented as the sum of two convolution operations: (1) a sparse sympathetic nervous system brain activity and a fast physiological smoothing system, (2) a periodic activation and a slow physiological smoothing system. The SC signal can be represented combining the phasic and tonic components as follows:
\begin{equation}
    y(t) = y_p(t) +y_s(t)
\end{equation}
where $y(t)$, $y_p(t)$ and $y_s(t)$ represent the SC signal, and the phasic and tonic components, respectively. The phasic component can be considered as a smoothed version of neural spiking activity from the brain. We model this smoothing filter using the first order diffusion kinetics of sweating from the sweat ducts to the strata cornea and subsequent evaporation from the strata cornea \cite{ans5}. We can combine both kinetics and use the following second order differential equation including the stimulation from sudomotor $u(t)$ to describe the phasic component in the sweat glands:

\begin{equation}
    \tau_{\gamma} \tau_{d} \frac{d^2 y_p(t)}{dt^2} + (\tau_{\gamma} + \tau_{d}) \frac{dy_p(t)}{dt} +y_p(t) = u(t)
\end{equation}

where $u(t)$, $\tau_{\gamma}$ and $\tau_{d}$ represent the neural stimuli generated by the sympathetic nervous system, the rise and decay times for each SC response, respectively. In order to obtain a solution to the system equation, we assume that initially the sweat duct is empty. Hence, the solution to the differential equation becomes,

\begin{equation}
    y_p(t) = y_p(0) \epsilon^{-\frac{t}{\tau_{\gamma}}} +h(t) * u(t)
\end{equation}
where $h(t)$ refers to the system response and can be represented as a scaled version of the Bateman function. Here, the operator '*' represents the convolution operation. $h(t)$ can be written as follows,

\begin{IEEEeqnarray*}{rCl}
h(t) &=&\left\{
\begin{aligned}
&\frac{1}   {\tau_\gamma - \tau_d}   \epsilon^{-\frac{t}{   \tau_\gamma  }} -  \epsilon^{ -\frac{t}{\tau_d}   } ;\;  &if\; t\ge 0 \\
&0 ;\;  &otherwise 
\end{aligned}\right\}\\
\end{IEEEeqnarray*}
Many SC decomposition strategies have been proposed by different researchers to recover the timings and amplitudes of neural stimulation as well as to estimate the underlying physiological parameters. We adopt Amit et. al. proposed block coordinate descent approach for skin conductance signal decomposition by employing generalized-cross-validation for balancing between smoothness of the tonic component, the sparsity of the neural stimuli, and residual error \cite{eda_feat}.

\begin{figure*}[!htb]
\begin{minipage}{0.35\textwidth}
\begin{center}
  \includegraphics[width=0.8\linewidth]{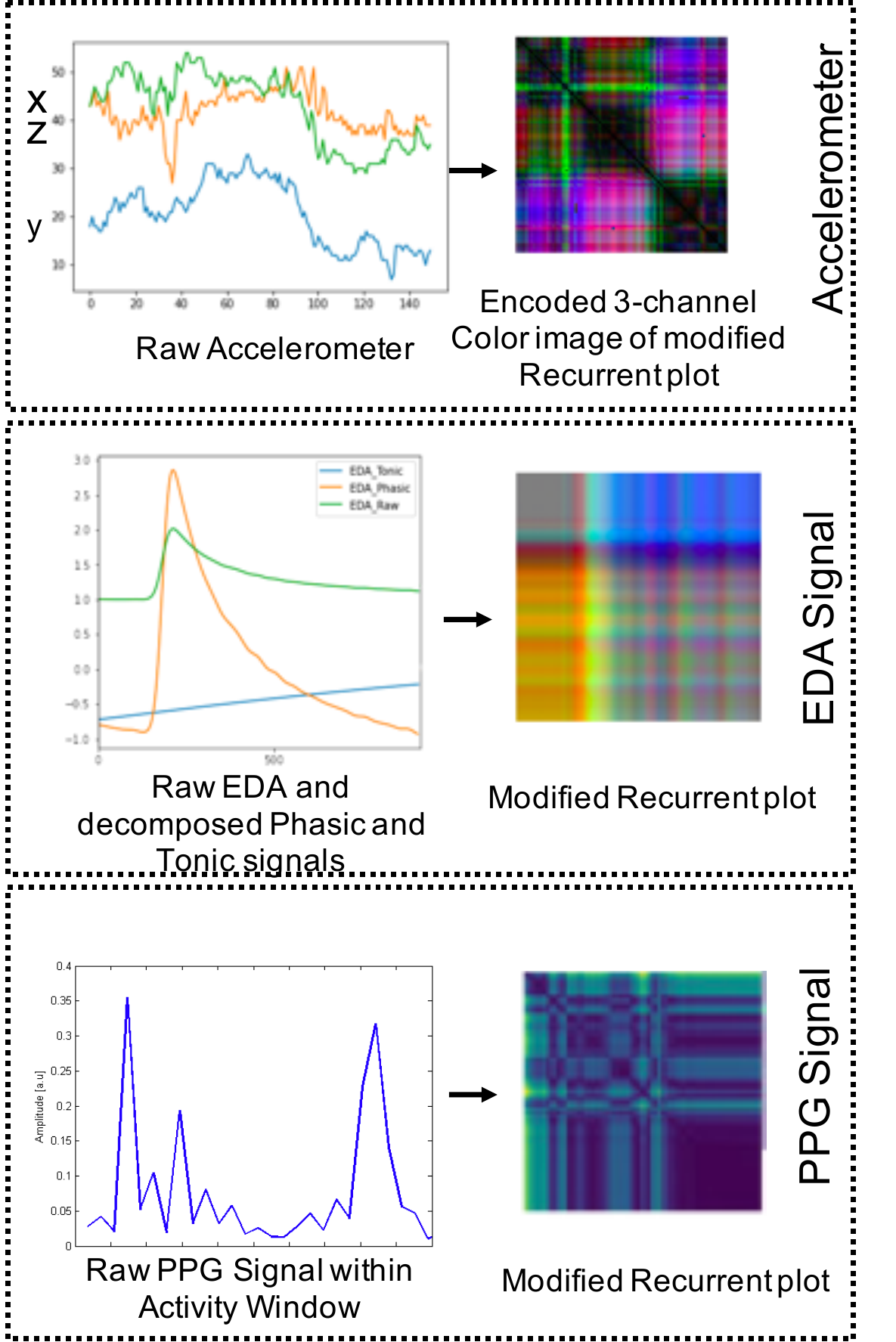}
  \caption{Modified Recurrence Plot (RP) encoding of 3-axis accelerometer signal into single 3-channel color image representation}
  \label{fig:acc_encode}
\end{center}
\end{minipage}
 \begin{minipage}{0.64\textwidth}
 \begin{center}
  \includegraphics[width=\linewidth]{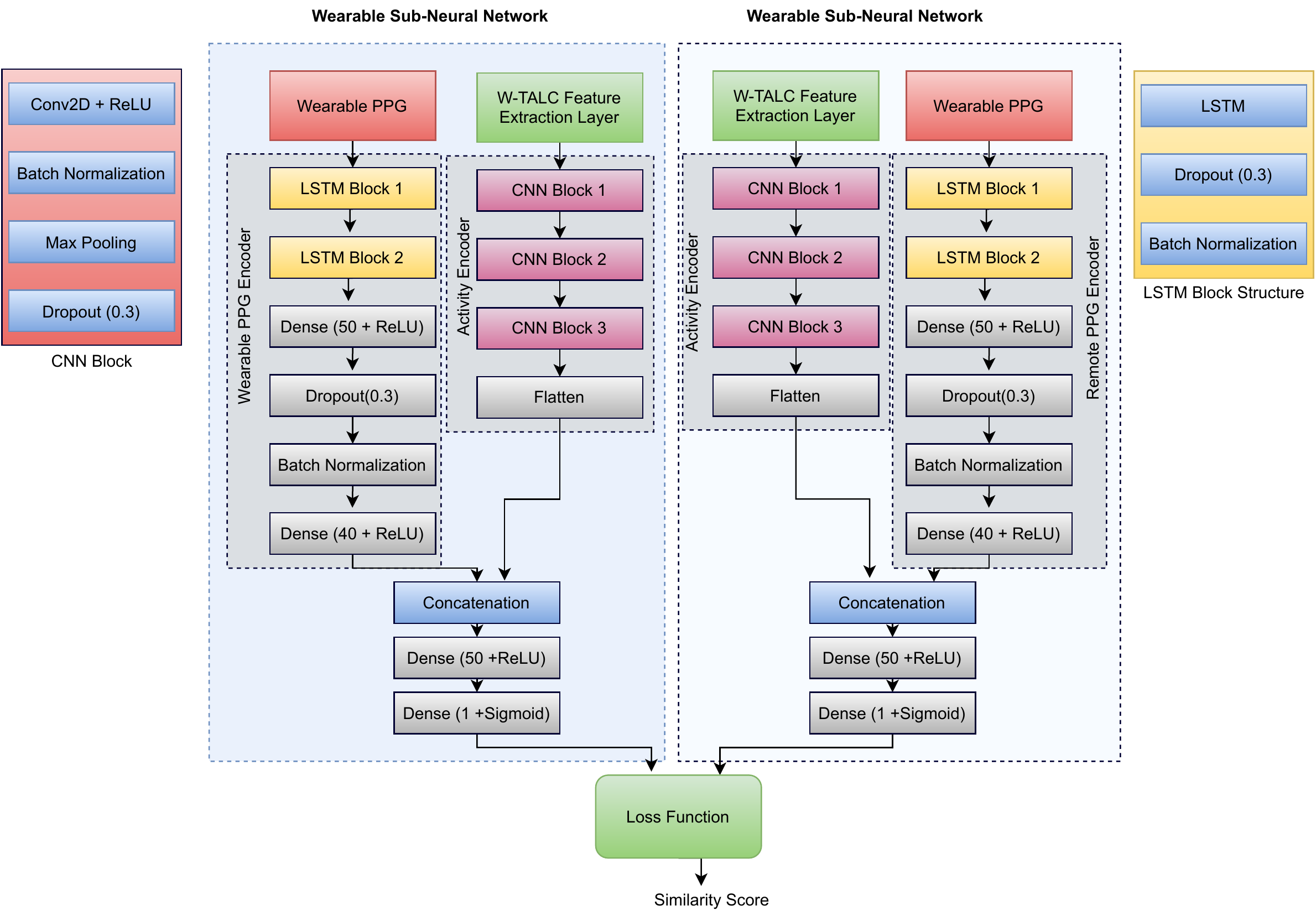}
  \caption{Schematic diagram of our proposed Multi-Modal Siamese Neural Network \emph{mmSNN} where the goal of learning the feature representation is to minimize the contrastive loss in case for corresponding data and maximize the contrastive loss otherwise.}
  \label{fig:mmSNN}
\end{center}
 \end{minipage}
\end{figure*}

\subsection{Image Representation of Sensor Data Within Gesture Window}
After localizing start and end point of a recognized gesture from hand worn accelerometer, we encode entire gesture window related accelerometer signal as images \cite{encode_acc} (Implementation Codes: \cite{encode_acc_code}). In this regard, as shown in Fig \ref{fig:acc_encode}, at first, we consider a Recurrence Plot (RP) of each accelerometer axis (3-axis) data. RP is a visualization tool for analyzing nonlinear data points on phase space trajectories of a dynamical system, whose states are typically in a rather intricate manner \cite{rp1}. We resort RP to encode 3-axis signals as RGB channels of color images so that their correlation information can be exploited. In this regard, we use Lu et. al. proposed modified RP encoding method to generate 3-channel color image from 3-axis accelerometer signal using a modified RP matrix construction method \cite{encode_acc} (Fig \ref{fig:acc_encode}). Similarly we can apply modified RP matrix construction method on 3-D raw, phasic and tonic signals of Electrodermal activity as welll as 1-D PPG signals as shown in Fig \ref{fig:acc_encode}.

\section{Multi-Modal Siamese Neural Network (mmSNN)}
Siamese Neural Network (SNN) consists of two similar sub-neural networks with identical parameters and shared weights \cite{siamese}. These two sub-neural networks are connected by a contrastive loss function, which calculates a similarity score between two input samples based on the feature vector representation of the two sub-neural networks. \emph{PhysioGait} extends SNN to Multi-Modal sub-neural networks where each sub-network consists of a CNN-based accelerometer image encoder and a LSTM-based physiological sensor (PPG/EDA/Temp) encoder. Both of the encoders (CNN and LSTM) are concatenated together to generate a generalized fused feature representation (Fig~\ref{fig:mmSNN}). Finally, two identical sub-networks are fused together using a weighted cost function.
\subsection{Multi-modal Feature Learning}
mmSNN takes a pair of inputs that consist of a accelerometer image and a time-series of physiological signal within a detected gesture window. To achieve maximized similarities between two embeddings, a Siamese Network was designed that (1) learns the joint embeddings between the physiological signal and accelerometer images using deep encoders and (2) maximizes the similarity measure between the two modalities. Let a multimodal dataset $X=\{ p_g,a_g \}$ where $p$ and $a$ represent $g^{th}$-gesture related physiological and accelerometer image. The objective is to build two encoders that convert the physiological signal and accelerometer image into a common space, $\eta$. The {\bf sequence encoder} i.e., physiological data encoder ($\phi$) maps the physiological signal into the common space was built using LSTM modules of different sizes to facilitate learning of temporal features from the physiological data. A linear layer was then applied to project the learned features into a one-dimensional feature vector in the common space, $\eta$. Similarly, the {\bf image encoder} i.e.  image representation of accelerometer encoder, $\alpha$, mapped the input image into the common space ($\eta$) using Convolutional Neural Networks (CNN) with different sizes of convolutional layers. Again, a linear layer was applied to produce a one-dimensional feature vector in the common space ($\eta$). The output of two encoders ($\phi$ and $\alpha$) was then concatenated using a compatibility function ($ F(p,a)=[ \phi (p), \alpha (a) ] $). $F$ is used to measure the similarity between two sample pair embeddings in the common space ($\eta$). Finally, the output of $F$ from two subnetworks is passed to a weighted loss function for joint identification and verification.
\subsection{Weighted Jointed Identification}
During the training process, we build on the joint identification and verification approach from \cite{41} to define our training objective. We use a softmax loss function to compute identification cost which has been integrated into our final training objective function defined as follows
\begin{equation}
V(\eta) = P(q=c|\eta) = \frac{exp(W_c\eta)}{\sum_k{exp(W_k\eta)}}
\end{equation}
where $\eta$ is the common feature vector out from multi-modal feature learning step and $q$ is the identity. $W_c$ and $W_k$ indicate the $c^{th}$ and the $k^{th}$ column of the softmax matrix $W$, respectively. $W$ is the feature weight matrix of common feature vector $\eta$. The weighted joint identification objective function, which is used for training, incorporates the ability to predict a persons identity.

\subsection{Image Encoder}
The image encoder ($\alpha$), that takes image representation of accelerometer sensor within activity window, consists of a series of convolution layers, where each convolution layer further followed by batch normalization, max pooling, and dropout layer. Here, we use a 2D CNN for extracting the spatial information from the input signature. The first convolutional layer takes the signature image (size: 155 x 220) as input and performs the filter operation with 8 filters of size 11 x 11. The outputs of the first convolutional layer are passed to the second convolutional layer (16 filters with a size of 5 x 5). Next, the outputs of the second convolutional layer are passed to the third convolutional layer (32 filters with a size of 3 x 3). Finally, the summary of all the spatial features of a signature image is passed to the flatten layer to produce a 1D feature vector.

\subsection{Sequence Encoder}
The sequence encoder ($\phi$) i.e. physiological signal encoder converts the physiological signals (PPG, EDA) into a common representation. To capture the longer dynamics in the temporal dimension of the physiological signal, we have used two consecutive LSTM layers. The physiological encoder consists of two LSTM layers; the first LSTM layer consists of 75 neurons, and the second one with 55 neurons. Each LSTM layer is followed by a dropout and batch normalization layer. Next, a dense layer of 50 neurons, followed by a dropout and batch normalization layer, is connected to another dense layer with 40 neurons. Informative features of the input physiological have been extracted using a series of LSTM layers. Each LSTM layer extracts significant temporal information from the physiological signal. LSTM layers are used to extracts deep temporal dynamics from the input data. A series of dense layers have been applied to make the network deeper, which leads to better performance.

\section{Experimental Evaluation}
In this section, we aim to evaluate our proposed \emph{PhysioGait} mode performance towards developing a personalized cognitive fatigue assessment system using wearables without any target labels.
\subsection{Datasets}
We use four datasets to evaluate \emph{PhysioGait} model performance which are described as follows:
\begin{itemize}

\item {\bf D1: Gamer's Fatigue Dataset}: We recruited 5 student video games players (age ranges from 19-25) for 7 days who stayed up during a 22 hour shift every alternative day (4 days each) to simulate cognitive fatigue while wearing Empatica E4 watch \cite{fatigue_data}. Empatica E4 watch consists of accelerometer (ACC), electrodermal activity (EDA), photoplethysmography (PPG) and skin temperature (TEMP) sensors. During the data collection (including non-gaming days), participates were asked to measure their sleepiness based on the `Stanford Sleepiness Scale' (SSS) \cite{sss,sss1}  (ranges 1-7 representing active to extremely sleepy) and the `Sleep-2-peak' score \cite{sleep2peak} (ranges 1-7 representing active to extremely sleepy) using Sleep2Peak Android App \cite{sleep2peakapp,s2papp}.

\item {\bf D2: Restaurant Data}: We recruited 8 student volunteers to perform a preparing and eating sandwich activity in restaurant environment for 20 minutes each while wearing Empatica E4 watch \cite{empatica}. Empatica E4 watch consists of accelerometer (ACC), electrodermal context (EDA), photoplethysmography (PPG) and skin temperature (TEMP) sensors.

\item {\bf D3: Older Adults Data}: This is prior collected data \cite{autocognisys} on 22 recruited older adults (75-95 years of old) who performed 13 scripted activities while wearing Empatica E4 watch \cite{empatica}.

\item {\bf D4: Healthy Adults Fatigue Dataset}: We have used publicly available health adults fatigue dataset \cite{luca20}. Data from 28 healthy individuals (26–55 years of age, average age 42 years, 41/51\% female/male), of which 17 enrolled up to 2 days after returning from long-haul flights with 3–7 time zone differences and hence were recovering from jet lag, from 1 to 219 consecutive days (mean 35, median 9, total 973 days) were collected. Objective data was collected using a multisensor wearable device, Everion (Biovotion AG, Switzerland \cite{biovotion}), in conjunction with a mobile app, SymTrack (Gastric GmbH, Switzerland), to deliver a daily fatigue questionnaire. Volunteers were asked to continuously wear the Everion device around their non-dominant arm over a 1-week period. The device combines a 3-axis accelerometer, barometer, galvanic skin response electrode, and temperature and photo sensors (see \cite{luca20} for more details).

\item {\bf D5: Combined Data of D1, D2 and D3}: We have merged D1, D2 and D3 datasets altogether as all of the datasets have used Empatica E4 with similar sensor data i.e. ACC, PPG, EDA and TEMP, thus total number of individuals becomes 35.

\item {\bf D6: Combined Data of D1, D2, D3 and D4}: We have merged D1, D2, D3 and D4 datasets altogether only for ACC, PPG and TEMP variables as D4 dataset has no EDA data thus total number of individuals becomes 63.

\end{itemize}
\subsection{Data Preparation}
To generate appropriate input to mmSNN, we need pairs of accelerometer and physiological signal windows that are similar, and, pairs of accelerometer and physiological signal windows that are dissimilar. The objective function $F$ will minimize the loss when similar pairs are found and maximize loss otherwise. The similar pairs will be associated with the user's identity in the training set. As stated in our {\bf hypotheses}, we consider our feature space vector $\eta_i$ will be similar to another $\eta_j$ if both belongs to same individuals given the same gesture, i.e., the gesture activity responses on physiological-gesture combined feature representation is a unique biometric for each individual. Considering the above hypothesis, first, we use pre-trained GestureKeeper \cite{gesture_keeper} model and used it to detect and localize gesture window. During the training of mmSNN, we consider user's identity of the training samples are known. We converted accelerometer signal to image and took physiological signal (sequential data) within the gesture window and train mmSNN model along with the identity function. Accelerometer image will ba passing through the accelerometer encoder and physiological signal will be passing through the physiological encoder while training mmSNN model for similarity/dissimilarity learning based user's identity classifier. We consider $accuracy = \frac{TP+TN}{TP+TN+FP+FN}$ and Standard Deviation $\pm\%$ as evaluation metrics for identifying users for all gesture window in the testing data. Since, our framework is the first of its kind, we could not compare with any other frameworks with our results.

\begin{figure}
  \centering
  \includegraphics[width=0.8\linewidth]{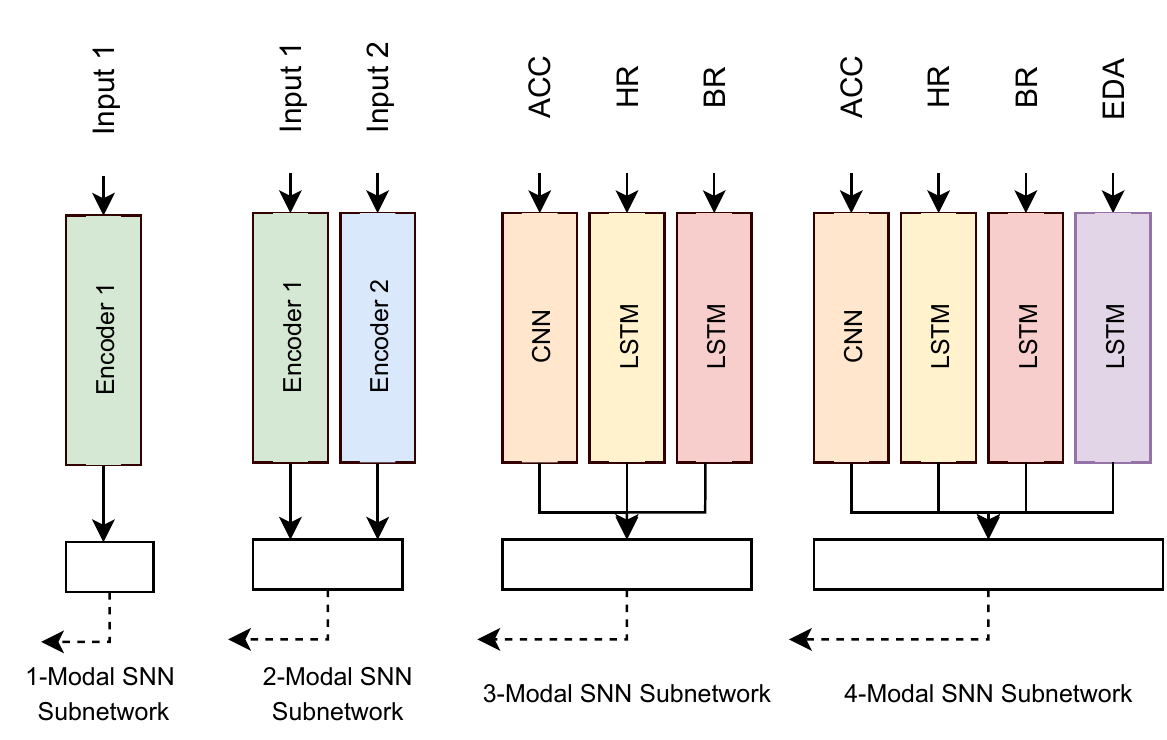}
  \caption{Core Architecture of Different Versions of PhysioGait model sub-network. This sub-network will be mirrored on the left to develop a contrastive triplet loss based Multi-Modal Siamese Network}
  \label{fig:MultiplePhysioGait}
\end{figure}

\subsection{Results}
We investigate different physiological signals i.e., Photoplethysmograph (PPG), Electrodermal Activity (EDA), Skin Temperature (Temp), EDA, Heart Rate (HR) derived from PPG signal \cite{hr_ppg}), Breathing Rate (BR) derived from PPG \cite{br_ppg}), BVP (Blood Volume Pulse derived from PPG), IBI (Inter Beat Interval derived from PPG), Tonic Component (TC) of EDA signal \cite{eda_feat}) and Phasic Component (PC) of EDA data \cite{eda_feat}) signals' privacy vulnerabilities using our proposed method.

\subsubsection{1-Modal Siamese Neural Network}
In this experiment, we consider one sub-network (1-Modal SNN) (Fig~\ref{fig:MultiplePhysioGait} 1-Modal SNN) 1) use cases: (1) CNN Encoder on each image representation of wearable signals' performance individually using our core CNN sub-network and (2) LSTM Encoder on each raw signal of wearables' performance individually using our core LSTM sub-network. Table \ref{tab:result_table_1} presents individually performed results of different sensing values that illustrates that BR and HR shows better performance than others individually for higher number of users' scenario (\#28). This table shows that accelerometer sensing performs better with CNN-based models while the others perform better in LSTM-based 1-modal SNN. It also can be depicted that temperature sensing does not provide enough promises in this regard.

\begin{table*}[!h]
  \begin{center}
\begin{small}
  \caption{Comparison of \emph{PhysioGait} performance on each dataset (D1, D2, D3, D4) of re-identifying user's identity using wearable based hand-gesture aware physiological sensor signals: Photoplethysmograph (PPG), Electrodermal Activity (EDA), Skin Temperature (Temp), EDA, Heart Rate (HR) derived from PPG signal \cite{hr_ppg}), Breathing Rate (BR) derived from PPG \cite{br_ppg}), BVP (Blood Volume Pulse derived from PPG), IBI (Inter Beat Interval derived from PPG), Tonic Component (TC) of EDA signal \cite{eda_feat}) and Phasic Component (PC) of EDA data \cite{eda_feat})}
  \label{tab:result_table_1}

  \begin{tabular}{|p{0.7cm}|p{1.45cm}|p{1.45cm}|p{1.45cm}|p{1.45cm}|p{1.45cm}|p{1.45cm}|p{1.45cm}|p{1.45cm}|p{1.45cm}}
      \hline
 \multicolumn{1}{c}{CNN} &  \multicolumn{1}{c}{Encoder} &  \multicolumn{1}{c}{Sub-network}&\multicolumn{1}{c}{ }&\multicolumn{1}{c}{ }&\multicolumn{1}{c}{ }&\multicolumn{1}{c}{ }&\multicolumn{1}{c}{ }&
    \\ \hline
 {\bf Data \#Users}     & {\bf PPG}  & {\bf HR} & {\bf BR} & {\bf BVP}& {\bf IBI}  & {\bf EDA } & {\bf ACC } & {\bf Temp}   \\
 \hline

    D1(5) & $69.65\pm5.4$  & $70.45\pm2.4$ & {\bf 71.58$\pm$3.1} & $66.65\pm4.6$  & $67.38\pm3.8$ & $52.43\pm9.4$ & $65.76\pm4.3$  & $49.23\pm8.2$    \\ \hline
    D2(8) & $67.12\pm5.1$  & $69.29\pm3.5$ & {\bf 70.54$\pm$2.9} & $65.86\pm3.9$  & $65.43\pm3.1$ & $50.87\pm9.1$ & $61.56\pm4.5$  & $48.43\pm7.2$    \\ \hline
    D3(22) & $66.44\pm6.6$  & {\bf 67.38$\pm$4.7} & $67.16\pm5.1$ & $64.95\pm4.5$  & $64.88\pm4.9$ & $50.63\pm9.6$ & $59.6\pm8.8$  &  $45.5\pm11.4$    \\ \hline
    D4(28) & $65.65\pm7.1$  & {\bf 66.95$\pm$4.3} & $66.85\pm4.1$ & $65.82\pm6.1$  & $66.75\pm5.3$ &  & $59.0\pm11.8$  &  $34.8\pm12.5$    \\ \hline
     D5(35) & $63.55\pm8.4$  & 62.54$\pm$5.8 & 64.65$\pm$4.1 & {\bf 64.75$\pm$7.3}  & $62.56\pm5.6$ & $50.43\pm9.5$  & $54.5\pm12.5$  &  $35.4\pm15.4$    \\ \hline
     D6(63) & $62.55\pm8.9$  & 61.76$\pm$5.8 & {\bf 63.77$\pm$5.2} & 63.52$\pm$6.8  & $62.45\pm5.5$ &  & $53.4\pm11.4$  &  $36.4\pm18.4$    \\ \hline

 \multicolumn{1}{c}{LSTM} &  \multicolumn{1}{c}{Encoder} &  \multicolumn{1}{c}{Sub-network}&\multicolumn{1}{c}{ }&\multicolumn{1}{c}{ }&\multicolumn{1}{c}{ }&\multicolumn{1}{c}{ }&\multicolumn{1}{c}{ }&
    \\ \hline
 {\bf Data \#Users}     & {\bf PPG}  & {\bf HR} & {\bf BR} & {\bf BVP}& {\bf IBI}  & {\bf EDA } & {\bf ACC } & {\bf Temp}   \\
 \hline
    D1(5) & $67.45\pm6.3$  & $65.45\pm3.5$ & {\bf 70.45$\pm$3.5} & $68.65\pm5.3$  & $65.45\pm3.4$ & $55.45\pm10$ & $60.52\pm5.5$  & $45.34\pm9.2$    \\ \hline
    D2(8) & {\bf 67.76$\pm$6.2}  & $65.56\pm4.2$ & {65.45$\pm$3.2} & $66.67\pm3.5$  & $64.56\pm4.3$ & $52.45\pm6.9$ & $60.25\pm5.2$  &  $51.55\pm6.3$    \\ \hline
    D3(22) & {\bf 66.45$\pm$7.2}  & {65.45$\pm$5.2} & $64.56\pm6.3$ & $66.45\pm5.1$  & $60.45\pm3.2$ & $51.35\pm8.2$ & $55.4\pm11.8$  &  $49.56\pm9.9$    \\ \hline
    D4(28) & $67.45\pm8.1$  & {\bf 68.45$\pm$4.4} & $67.5\pm11.2$ & $64.45\pm5$  & $67.43\pm4.8$ &  & $52.9\pm12.8$  &  $35.4\pm11.4$    \\ \hline
     D5(35) & $66.67\pm7.9$  & {\bf 68.54$\pm$4.6} & 66.65$\pm$3.6 & 64.37$\pm$8.8  & $66.54\pm6.5$ & $51.56\pm8.4$  & $52.8\pm10.5$  &  $32.4\pm20.1$    \\ \hline
     D6(63) & $65.36\pm7.8$  & {\bf 67.54$\pm$5.1} & 65.35$\pm$4.8 & 64.56$\pm$7.2  & $64.34\pm5.1$ &  & $50.4\pm11.4$  &  $33.3\pm15.3$    \\ \hline
  \end{tabular}
\end{small}
  \end{center}
\end{table*}

\subsubsection{2-Modal Siamese Neural Network}
We consider two sub-networks in the core Siamese Neural Network architecture (2-Modal SNN) (Fig~\ref{fig:MultiplePhysioGait} 2-Modal SNN) under the following use cases (1) Encoder 1: CNN, Encoder 2: CNN, (2) Encoder 1: CNN, Encoder 2: LSTM, (3) Encoder 1: LSTM, Encoder 2: LSTM. Table \ref{tab:result_table_2} presents the detail results of different combination of the above 2-modal SNN architectures. It can be easily seen that the 2-modal SNN works better when we place accelerometer in encoder 1 as CNN-subnetwork and any other sensing to encoder 2 as LSTM-subnetwork.

\begin{table*}[!h]
  \begin{center}
\begin{small}
  \caption{Comparison of \emph{PhysioGait} performance on each dataset (D1, D2, D3, D4) of re-identifying user's identity using wearable (Top 9 performing combinations are presented)}
  \label{tab:result_table_2}

  \begin{tabular}{|p{0.7cm}|p{1.45cm}|p{1.45cm}|p{1.45cm}|p{1.45cm}|p{1.45cm}|p{1.45cm}|p{1.45cm}|p{1.45cm}|p{1.45cm}|p{1.45cm}|}
      \hline
  \multicolumn{1}{c}{Encoder} &  \multicolumn{1}{c}{1} &  \multicolumn{1}{c}{CNN}&\multicolumn{1}{c}{and} &\multicolumn{1}{c}{Encoder} &  \multicolumn{2}{c}{2} &  \multicolumn{1}{c}{CNN}&\multicolumn{1}{c}{ }&
    \\ \hline
 {\bf Data \#Users}     & {\bf ACC \& PPG}  & {\bf ACC \& HR} & {\bf ACC \& BR} & {\bf ACC \& BVP}& {\bf ACC \& IBI}  & {\bf ACC \& EDA } & {\bf PPG \& EDA } & {\bf HR \& EDA } & {\bf BVP \& EDA}   \\
 \hline

    D1(5) & $86.56\pm2.4$  & $87.45\pm2.7$ & {\bf 90.5$\pm$1.0} & $86.56\pm2.1$  & $80.56\pm2.6$ & $70.45\pm7.4$ & $67.54\pm3.5$  & $66.45\pm7.2$ & $60.2\pm9.9$    \\ \hline
    D2(8) & $82.54\pm4.3$  & {\bf 86.38$\pm$2.7} & $86.20\pm2.5$ & $83.56\pm3.0$  & $75.58\pm2.5$ & $66.67\pm7.2$ & $66.84\pm7.3$  & $66.85\pm7.2$ & $54.7\pm10.5$    \\ \hline
    D3(22) & $78.46\pm5.3$  & {\bf 80.56$\pm$3.8} & $79.95\pm3.2$ & $75.32\pm4.9$  & $70.54\pm4.3$ & $58.45\pm8.5$ & $59.4\pm10.5$  & $59.8\pm11.4$ & $50.4\pm12.5$    \\ \hline
    D4(28) & $75.45\pm4.8$  & {\bf 79.56$\pm$2.8} & $79.05\pm2.4$ & $74.67\pm4.0$  & $70.25\pm3.2$ & &   &  &  \\ \hline
     D5(35) & $75.54\pm5.5$  & {\bf 76.45$\pm$5.1} & $76.45\pm4.5$ & $72.23\pm5.1$  & $66.75\pm3.9$ & $56.34\pm7.5$ & $56.34\pm8.2$  & $56.43\pm7.3$ & $50.45\pm9.9$    \\ \hline
     D6(63) & {\bf  75.56$\pm$4.9}  & 73.56$\pm$3.8 & $75.11\pm5.3$ & $71.45\pm3.3$  & $66.67\pm4.2$ && & &    \\ \hline
    
  \multicolumn{1}{c}{Encoder} &  \multicolumn{1}{c}{1} &  \multicolumn{1}{c}{LSTM}&\multicolumn{1}{c}{and} &\multicolumn{1}{c}{Encoder} &  \multicolumn{2}{c}{2} &  \multicolumn{1}{c}{LSTM}&\multicolumn{1}{c}{ }&
    \\ \hline
 {\bf Data \#Users}     & {\bf ACC \& PPG}  & {\bf ACC \& HR} & {\bf ACC \& BR} & {\bf ACC \& BVP}& {\bf ACC \& IBI}  & {\bf ACC \& EDA } & {\bf PPG \& EDA } & {\bf HR \& EDA } & {\bf BVP \& EDA}   \\
 \hline
    D1(5) & $82.56\pm3.2$  & $87.43\pm2.8$ & {\bf 87.95$\pm$2.6} & $83.76\pm2.4$  & $80.48\pm4.3$ & $76.56\pm5.1$ & $70.54\pm3.2$  & $76.65\pm6.7$ & $73.56\pm5.6$    \\ \hline
    D1(8) & $80.55\pm4.5$  & {\bf 84.34$\pm$3.2} & 82.45$\pm$3.9 & $81.56\pm3.3$  & $79.45\pm5.5$ & $75.45\pm5.9$ & $70.56\pm4.3$  & $73.45\pm6.0$ & $70.65\pm5.1$    \\ \hline
    D3(22) & $75.45\pm5.8$  & {\bf 80.42$\pm$4.3} & 78.45$\pm$3.2 & $78.53\pm4.3$  & $75.45\pm4.3$ & $69.56\pm5.1$ & $68.45\pm5.9$  & $70.76\pm6.4$ & $65.76\pm6.3$    \\ \hline
    D4(28) & $74.56\pm6.4$  & 76.54$\pm$5.7 & {\bf 77.57$\pm$5.1} & $75.56\pm4.5$  & $73.55\pm5.6$ &  &   & &    \\ \hline
     D5(35) & $71.65\pm3.6$  & {\bf 75.63$\pm$6.8} & $75.42\pm7.1$ & $72.63\pm6.7$  & $66.53\pm8.3$ & $55.56\pm9.2$ & $56.45\pm7.4$  & $56.88\pm8.2$ & $51.5\pm10.5$    \\ \hline
     D6(63) & {\bf  72.95$\pm$5.3}  & 72.55$\pm$4.3 & $71.56\pm6.3$ & $72.84\pm2.2$  & $69.85\pm6.2$ && & &    \\ \hline
     
  \multicolumn{1}{c}{Encoder} &  \multicolumn{1}{c}{1} &  \multicolumn{1}{c}{CNN}&\multicolumn{1}{c}{and} &\multicolumn{1}{c}{Encoder} &  \multicolumn{2}{c}{2} &  \multicolumn{1}{c}{LSTM}&\multicolumn{1}{c}{ }&
    \\ \hline
 {\bf Data \#Users}     & {\bf ACC \& PPG}  & {\bf ACC \& HR} & {\bf ACC \& BR} & {\bf ACC \& BVP}& {\bf ACC \& IBI}  & {\bf ACC \& EDA } & {\bf BR \& ACC } & {\bf HR \& ACC } & {\bf BVP \& ACC}   \\
 \hline
    D1(5) & $86.56\pm2.4$  & $87.45\pm2.7$ & {\bf 90.5$\pm$1.0} & $86.56\pm2.1$  & $80.56\pm2.6$ & $70.45\pm7.4$ & $67.54\pm3.5$  & $66.45\pm7.2$ & $60.2\pm9.9$    \\ \hline
    D2(8) & $82.54\pm4.3$  & {\bf 86.38$\pm$2.7} & $86.20\pm2.5$ & $83.56\pm3.0$  & $75.58\pm2.5$ & $66.67\pm7.2$ & $66.84\pm7.3$  & $66.85\pm7.2$ & $54.7\pm10.5$    \\ \hline
    D3(22) & $78.46\pm5.3$  & {\bf 80.56$\pm$3.8} & $79.95\pm3.2$ & $75.32\pm4.9$  & $72.54\pm4.3$ & $58.45\pm8.5$ & $59.4\pm10.5$  & $59.8\pm11.4$ & $50.4\pm12.5$    \\ \hline
    D4(28) & $77.45\pm4.8$  & {\bf 79.56$\pm$2.8} & $79.05\pm2.4$ & $75.67\pm4.0$  & $71.25\pm3.2$ & $55.45\pm8.9$ & $53.4\pm11.5$  & $60.0\pm11.8$ & $53.4\pm12.5$    \\ \hline
     D5(35) & $77.01\pm6.8$  & {\bf 78.75$\pm$6.2} & $78.45\pm4.3$ & $75.05\pm6.2$  & $70.54\pm5.7$ & $55.11\pm9.3$ & $55.85\pm5.6$  & $61.84\pm8.2$ & $52.84\pm7.5$    \\ \hline
     D6(63) & {\bf  77.00$\pm$3.8}  & 77.63$\pm$4.6 & $78.45\pm5.5$ & $74.65\pm5.4$  & $70.54\pm5.3$ && & &    \\ \hline
  \end{tabular}
\end{small}
  \end{center}
\end{table*}

\subsection{Baseline Models}
We carefully designed two multimodal SNN models that perform better than any other combinations of multimodal SNN framework (shown in 
Fig~\ref{fig:MultiplePhysioGait}): (1) 3-modal SNN that consists of Acceleromter-CNN encoder, HR LSTM encoder and BR LSTM encoder; and (2) 4-modal SNN that consists of accelerometer-CNN encoder, HR LSTM encoder, BR LSTM encoder and EDA LSTM encoder. We implement 4 state-of-art wearable-based privacy attack models to compare performances with \emph{PhysioGait} attack model. We also developed 6 versions of our mmSNN models to develop similar use-cases as 4 selected state-of-art models to solidify our contributions.

To evaluate the performances of \emph{PhysioGait} attack model, we develop following baseline models.
\begin{itemize}
\item {\bf B1 (BioInsights with accelerometer only):} This model focused on wearable (watch and google glass) accelerometer and gyroscope based 3 body-posture (pre-post exercise) classification and personalized biometric signature identification using Support Vector Machine algorithm \cite{wear_privacy1}. Due to the unavailability of both accelerometer and gyroscope in our datasets, we consider only accelerometer signals on BioInsights model as our baseline.

\item {\bf B2 (Benegui Model with accelerometer only):} This model proposed a gray-scale image representation of wearable sensor and developed a Convolutional Neural Network (CNN) based few shot user identification technique \cite{wear_privacy3}.

\item {\bf B3 (Kim Model):} This framework proposed bidirectional long short-term memory (LSTM)-based deep recurrent neural networks (DRNN) through late-fusion to develop a real-time system for ECG-based biometrics identification and classification \cite{ecg_privacy2}. We applied a similar method on PPG extracted HR in our datasets.

\item {\bf B4 (Ko Model):} This framework proposed an adjusted (Q,S) algorithm that automatically balance the R-peak distribution of PPG signal extracted HR towards modeling biometric signature for person re-identification from PPG data \cite{ecg_privacy3}.

\item {\bf P1:} 1-modal SNN with image representation of accelerometer sensor and CNN encoder to develop as similar use-case as B1 and B2 baseline models.

\item {\bf P2:} 1-modal SNN with image representation of PPG sensor signal and LSTM encoder to develop as similar use-case as B3 and B4 baseline models.

\item {\bf P3:} 3-modal SNN arranged as follows in each Siamese Sub-network, accelerometer image CNN encoder, HR LSTM encoder and BR LSTM encoder.

\item {\bf P4:} 4-modal SNN arranged as follows in each Siamese Sub-network, accelerometer image CNN encoder, HR LSTM encoder, BR LSTM encoder and electrodermal activity LSTM encoder.

\end{itemize}

\begin{table*}[!h]
  \begin{center}
\begin{small}
  \caption{Accuracy comparisons among different versions of mmSNN and state-of-art privacy attack frameworks on wearables}
  \label{tab:result_table_3}

  \begin{tabular}{|p{0.8cm}|p{1.45cm}|p{1.45cm}|p{1.45cm}|p{1.45cm}|p{1.45cm}|p{1.45cm}|p{1.45cm}|p{1.45cm}|p{1.45cm}}
  
    \hline
 {\bf Data \#Users}     & {\bf B1}   & {\bf B2} & {\bf P1} & {\bf B3} & {\bf B4} & {\bf P2 } & {\bf P3 } & {\bf P4}   \\
 \hline
     D5(35) & $75.76\pm5.2$   & 77.54$\pm$3.2 &  {\bf 79.56$\pm$2.2}  & 82.53$\pm$6.1 & 84.76$\pm$2.5& {\bf 88.43$\pm$4.1} & {\bf 90.56$\pm$0.7}  & {\bf 93.45$\pm$0.1}     \\ \hline
     
     D6(63) & $74.36\pm4.7$   & 75.43$\pm$2.5 & {\bf 78.56$\pm$3.3} & 79.56$\pm$4.8 & 81.54$\pm$7.4   & {\bf 82.77$\pm$2.8}  &  {\bf 89.45$\pm$0.2} &   \\ \hline

  \end{tabular}
\end{small}
  \end{center}
\end{table*}

\subsection{Results}
\begin{figure}
  \centering
  \includegraphics[width=\linewidth]{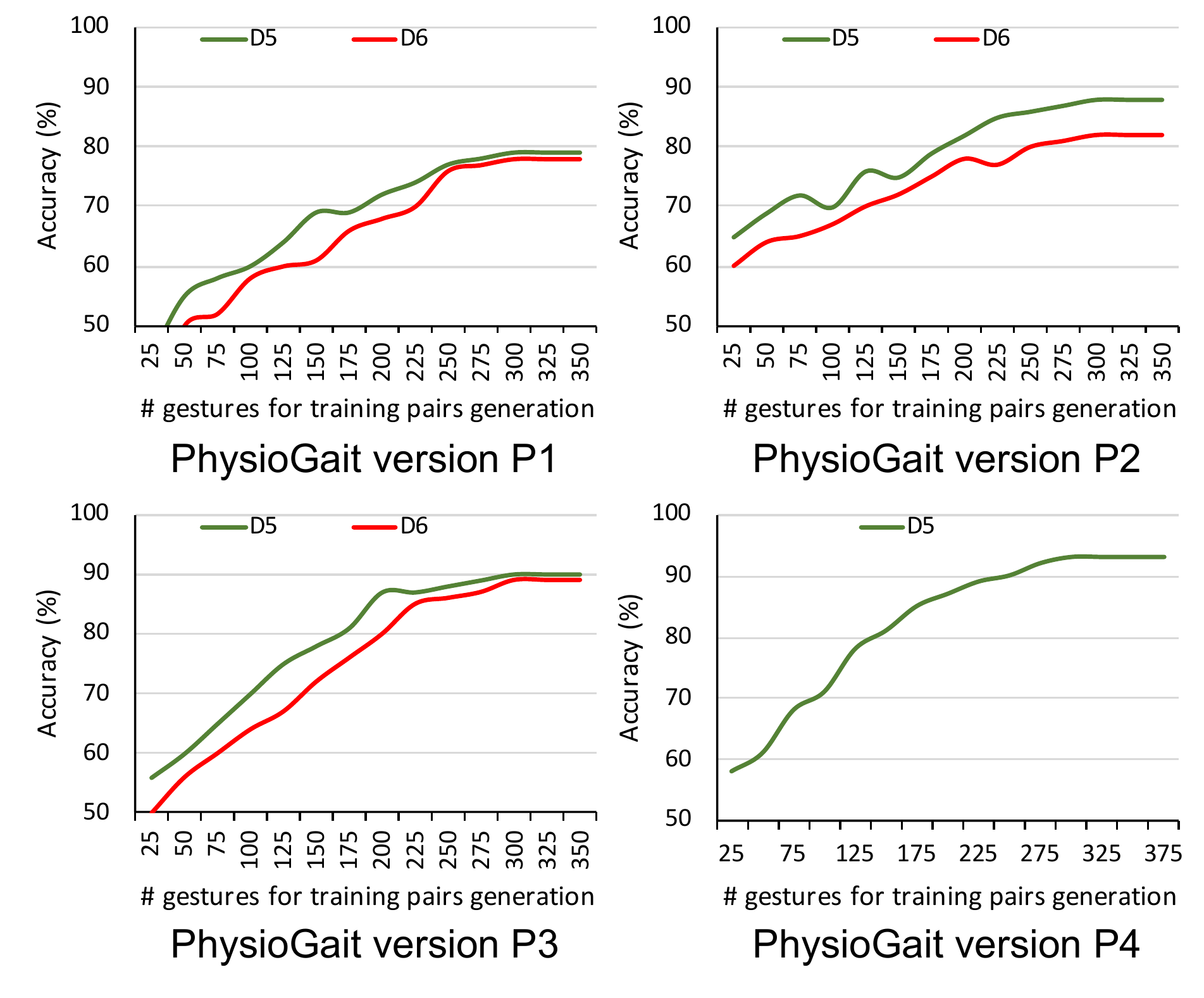}
  \caption{User's re-identification accuracy changes over number of gesture windows in the training sample for gesture aware PPG signal}
  \label{fig:results_gesture_attack}
\end{figure}

Table \ref{tab:result_table_3} presents the detailed results of our baseline attack models (B1, B2, B3 and B4) and different versions of \emph{PhysioGait} attack models (P1, P2, P3 and P4). To generate these results, we randomly selected 300 episodes of detected gestures as our training candidates which have been used to generate similar and dissimilar data pairs to train our core mmSNN model. The results show that different versions of our proposed \emph{PhysioGait} attack models outperform their similar state-of-art baseline models i.e. P1 outperforms B1 and B2, P2 outperforms B3 and B4. Moreover, we can see that our novel fusion models (P3 and P4) significantly outperform any other existing state-of-art model for both datasets D5 (with 35 individuals) and D6 (with 63 individuals). The number 300 episodes of detected gestures has been chosen based on our experiment on D1, D2, D3, D4, D5 and D6 datasets as shown in Fig~\ref{fig:results_gesture_attack} where we can see that after 300 episodes, the accuracies of person re-identification become stable (converged) for all of our baseline models P1, P2, P3 and P4.

\section{Conclusion}
The evolution of ubiquitous sensing technologies has enabled modern intelligent systems to monitor physical (such as cooking, walking, gestures) and physiological (such as heart rate, breathing rate) contexts more accurately than ever in privacy preserving way. Because wearable and ambient sensing technologies in conjunction with complex deep learning models are getting more powerful in detecting microscoping human contexts, data privacy has become a serious concern given the easy availability of multiple technologies. The Health Insurance Portability and Accountability Act (HIPAA) \cite{hipaa} states no clear indication of storing de-identified multi-source wearable sensor-based health contextual data in a single location (HIPAA compliant server). In this paper, we argue that, contextual responses of multi-source healthcare data can easily be utilized to perform a person re-identification attack if the attacker can learn few gesture related physiological features of the individual. While modern computer vision technology can be easily utilized to learn hand gestures and corresponding physiological signals (heart rate, breathing rate) from public surveillance camera, these huge amount of recorded videos can be easily utilized by the attackers to learn user specific biometrics to reveal identity from HIPPA compliant server stored wearable sensing data.

\end{document}